# Spin-polarized reflection of electrons in a two-dimensional electron system

Hong Chen, J. J. Heremans*, J. A. Peters, J. P. Dulka, and
A. O. Govorov,

Department of Physics and Astronomy, and The Nanoscale and Quantum Phenomena Institute, Ohio University, Athens OH 45701

N. Goel, S. J. Chung, and M. B. Santos

Department of Physics and Astronomy, and Center for Semiconductor Physics in Nanostructures, The University of Oklahoma, Norman OK 73019

Abstract

We present a method to create spin-polarized beams of ballistic electrons in a two-dimensional electron system in the presence of spin-orbit interaction. Scattering of a spin-unpolarized injected beam from a lithographic barrier leads to the creation of two fully spin-polarized side beams, in addition to an unpolarized specularly reflected beam. Experimental magnetotransport data on InSb/InAlSb heterostructures demonstrate the spin-polarized reflection in a mesoscopic geometry, and confirm our theoretical predictions.

*Corresponding author: Department of Physics and Astronomy, Clippinger Laboratories, Ohio University, Athens OH 45701 (heremans@ohiou.edu)

The spin of electrons and holes in semiconductor heterostructures has attracted much interest, as a factor to realize new spin-based electronic device concepts [1], and for its potential in realizing quantum computational schemes [2]. In heterostructures, spin can manifest itself through strong and tunable spin-orbit interaction terms [3]. Recent studies have often regarded spin-orbit interaction (SOI) as deleterious, since it can lead to short spin-coherence times. However, semiconductor heterostructures can be fabricated with a long carrier mean free path, longer than lateral dimensions within reach of present lithographic techniques. If the mean free path is longer than the lateral dimensions, charge transport in the geometry occurs ballistically, *i.e.* the preponderant scattering events involve the device boundaries [4]. In such mesoscopic devices, the decoherence due to SOI is minimized, and SOI, together with the device geometry, can be exploited for spin manipulation, and for the preparation of spin-polarized carrier states. Theoretical studies have explored the effect of SOI on one-dimensional mesoscopic transport, and on vertical transport through heterostructures [5]. Here we present a method to create spin-polarized beams of ballistic electrons by utilizing elastic scattering off a barrier in a straightforward open geometry, and present experimental results verifying the realization of the method. As illustrated in the upper panels of Fig. 1, a beam of two-dimensional electrons in a heterostructure is injected towards a barrier. Both energy and the momentum parallel to the barrier are conserved during the scattering event off the barrier. However, in the presence of SOI, scattering off the barrier leads to spin-flip events, and results in different reflection angles for different spin polarizations. The spin-polarized reflected beams can then be captured through suitably positioned apertures (upper left panel in Fig. 1). The multi-beam reflection process can be utilized to create spin-polarized electron populations, without the use of ferromagnetic contacts [6], and toward the preparation of specific spin states for quantum computational purposes.

The lower left panel in Fig. 1 shows the sample geometry. Equilateral triangles of inside dimensions 3.0 μm feature apertures, of conducting widths of ~ 0.2 μm, on two sides, while the left side forms the scattering barrier. Several triangles are measured in parallel [7]. The triangles were wet etched into *n*-type InSb/InAlSb heterostructures after electron beam lithography. The gentle wet-etching procedure affords highly reflecting barriers in III-V heterostructures [4]. Carriers enter the geometry from the top, travel ballistically to the left barrier, reflect off the latter, and exit through the bottom aperture. The total distance, including the reflection, between the apertures, amounts to 2.6 μm. The heterostructures were grown by molecular beam epitaxy on GaAs substrates, and consisted of a 20 nm wide InSb well, where the two-dimensional electron system (2DES) resides, flanked by $In_{0.91}Al_{0.09}Sb$ barrier layers [8]. Electrons are provided by Si $\delta$-doped layers on both sides of the well, separated from the 2DES in the well by 30 nm spacers. A third Si doped layer lies close to the heterostructure surface. All measurements were performed at 0.5 K, and at this temperature, a density $N_S = 2.6 \times 10^{11}$ cm$^{-2}$ and a mobility of 150,000 cm$^2$/Vs provide a mean free path of ~ 1.3 μm. Although shorter than the distance between the two apertures in Fig. 1, this mean free path is sufficiently long to ensure observation of a signal due to a ballistic trajectory. Indeed, the cutoff of the signal at the mean free path is not abrupt, but rather is characterized by a gradual decay of the signal amplitude [4,7]. The InSb well material features a narrow energy gap, a small effective mass, and also a strong SOI. Two SOI mechanisms can lead to the spin-dependent reflection effect: the Bychkov-Rashba mechanism, originating in the inversion asymmetry of the 2DES confining potential, and the Dresselhaus mechanism, from the bulk inversion asymmetry [3]. Experimental values for the SOI parameters in InSb-based heterostructures have only recently been accessed by optical measurements, and the preliminary data confirms SOIs larger than in most other III-V materials [9].

In our measurements, a current is drawn between the two apertures (Fig. 1), and the resulting voltage drop is measured as a function of a magnetic field applied perpendicular to the plane of the 2DES. In the semi-classical limit, the magnetic field $B$ serves to slightly deflect the ballistic carriers from linear trajectories, and thus to sweep the trajectories over the exit aperture. As an illustration of the role of $B$, the upper left panel of Fig. 1 shows trajectories calculated for our sample geometry and electron density (details of the calculation follow). The interaction with the barrier gives rise to three reflection angles, and the exit (lower) aperture is sufficiently wide to accommodate the three resulting exiting beams. For a narrow range of $B$, all three beams can be aimed to pass through the aperture. Varying $B$ in either direction causes the beams to be sequentially cut off, either by one side of the aperture or by the other. Each cutoff results in a stepwise rise in the potential or resistance measured over the structure. We expect 3 cutoffs on each side of the sweep, or 6 steps in total. In the experimental realization, the cutoffs will not be sharp, since we expect an angular spread at the injection aperture, albeit narrowed by the previously observed collimating effect of mesoscopic apertures [10]. Figure 2 contains experimental data for two separate samples ($S_1$ and $S_2$), plotted as the four-contact resistance measured over the triangular structures, versus applied $B$. For sample $S_1$, 6 minima appear at low $B$, superimposed on a negative magnetoresistance weak-localization background. The 6 minima are interpreted to result

Hong Chen *et al.*	Page 4 of 14	8/27/2003

from the stepwise increase in resistance as $B$ is varied, added to the negative magnetoresistance background. We also note here that the wet-etching process results in uncertainty in the structure's dimensions, and that therefore a non-zero $B$ may have to be applied to center the three beams on the exit aperture. Hence, the 6 minima need not be centered around $B = 0$. Sample $S_2$ underwent a deeper wet-etch, resulting in narrower apertures, as betrayed by the higher resistance values. Hence, the range of $B$ where three beams fit into the exit aperture of $S_2$ is reduced as compared to $S_1$. Two steps in resistance occur in such a narrow range of $B$ that they are observed as one, resulting in 5 observable minima. Assuming that the Bychkov-Rashba mechanism leads to the observed minima, the data can be used to estimate the magnitude of the spin splitting. SOI can be evaluated by the spin-splitting $\Delta_{SO}$ at the Fermi level $E_F$, given by $\Delta_{SO} = 2\alpha_{SO} k_F$, where $k_F$ denotes the Fermi wave vector and $\alpha_{SO}$ depends on material and heterostructure parameters. Estimating $\alpha_{SO}$ from the experiments, we have calculated the values of $B$ where cutoffs occur, using the equations derived below for the angular deviations from specular reflection, $\Delta\theta_{+\rightarrow-}$ and $\Delta\theta_{-\rightarrow+}$ (Fig. 1). The following parameters are consistent with our experimental observations: $\alpha_{SO} \approx 1\times 10^{-6}$ m$e$V cm and $\Delta_{SO} \approx 2.5$ m$e$V, at $N_S = 2.6\times 10^{11}$ cm$^{-2}$ and $E_F = 35$ m$e$V (the effective mass $m_e = 0.014 m_0$). This value for $\Delta_{SO}$ approaches that obtained from the optical measurements on similar InSb/InAlSb heterostructures [9]. The literature does not yet contain experimental values for the Dresselhaus SOI parameters in InSb-based heterostructures. Returning to the negative magnetoresistance background, we have consistently observed only a weak-localization peak at $B \approx 0$ in mesoscopic geometries fabricated in the InSb/InAlSb heterostructure, in contrast to the antilocalization signature observed in GaAs or InAs based 2DESs [11]. Another example of a weak-localization peak in a mesoscopic geometry is shown in the inset in Fig. 2, namely the resistance *vs.* applied perpendicular $B$, measured over an anti-dot lattice fabricated on the same heterostructure [12]. The absence of antilocalization is not surprising in InSb. Antilocalization requires the Dyakonov-Perel' spin scattering mechanism to dominate, leading to a randomization of the spin precession process due to a weak SOI [11]. Yet, due to large spin splitting in InSb, the impurity broadening of the electron energy is less than the spin-splitting, invalidating the conditions for Dyakonov-Perel' scattering and antilocalization ($\hbar/\tau \approx 0.5$ m$e$V $<< \Delta_{SO} \approx 2.5$ m$e$V, where $\tau$ is the scattering time deduced from the mobility mean free path).

We now present the theoretical description of spin-polarized multi-beam reflection utilized to generate the trajectories in Fig. 1. The motion of single electrons in a 2DES with SOI is described by the Hamiltonian: $\hat{H} = (\hat{p}_x^2 + \hat{p}_y^2)/2m_e + U(x) + \hat{V}_{SO}(x)$, where $\hat{p}_{x(y)}$ are the in-plane momentum operators, $U(x)$ is the lateral potential describing the reflection barrier and $\hat{V}_{SO}(x)$ represents the SOI operator. For the SOI, we assume the Bychkov-Rashba inversion asymmetry mechanism engendered by the electric fields perpendicular to, and in the plane of, the 2DES [3]. The SOI-operator is composed of two terms,

$$\hat{V}_{SO}(x) = \frac{\alpha_{SO}}{\hbar}(\hat{\sigma}_x \hat{p}_y - \hat{\sigma}_y \hat{p}_x) + \frac{\gamma}{\hbar}\frac{dU(x)}{dx}\hat{\sigma}_z \hat{p}_y, \qquad (1)$$

where $\hat{\sigma}_{x(y)}$ represent the Pauli matrixes. The first term in Eq. 1 originates from the perpendicular electric field $F_z$, the second from the in-plane electric field. The material parameter $\gamma$ describes the strength of the SOI, and $\alpha_{SO} = -e\gamma F_z$. The operator (1) assumes averaging in the z-direction over the wave function in the well.

Since the potential in the Hamiltonian depends only on the x-coordinate, the general solution of the Schrödinger equation takes a form $\Psi = \Phi(x)e^{ik_y y}$, where $k_y$ is the y-component of the momentum. Outside of the interaction zone with the barrier the wave function and energy of a single incident electron, have a form:

$$\Psi^{in}_{\mathbf{k},\pm} = \frac{1}{\sqrt{2}}\begin{pmatrix} 1 \\ \pm e^{i\varphi(\mathbf{k})}/i \end{pmatrix} e^{i\mathbf{kr}}, \quad E_{\mathbf{k}\pm} = \frac{\hbar^2 k^2}{2m_e} \pm \alpha_{SO} k, \qquad (2)$$

where $\mathbf{k} = (k_x, k_y)$ and $\tan\varphi(\mathbf{k}) = k_y/k_x$. For the above spin states (+ or -), the spins are perpendicular to the momentum due to the SOI. During the reflection process, the electron conserves both energy and $k_y$, leading to: $\Psi^{out} = A(+)\Psi_{\mathbf{q}_+,+} + A(-)\Psi_{\mathbf{q}_-,-}$, where the momenta of the reflected waves, $\mathbf{q}_+ = (q_{x+}, k_y)$ and $\mathbf{q}_- = (q_{x-}, k_y)$, are determined by kinematics equations. If the incident electron is in the state $\Psi^{in}_{\mathbf{k},+}$, the momentum $q_{x+} = -k_x$, while $q_{x-} = -k_x + \delta q_{x-}$ is determined by conservation of energy: $E_+(k_x, k_y) = E_-(-k_x + \delta q_{x-}, k_y)$. In the case of incoming state $\Psi^{in}_{\mathbf{k},-}$, the momenta of the reflected waves will be $q_{x-} = -k_x$, and $q_{x+} = -k_x + \delta q_{x+}$. From the above consideration, we deduce that the reflected wave for each incident state is composed of two beams propagating at different angles: for $\Psi_{k,+}^{in}$ at $\theta$ and $\theta_+$, and for $\Psi_{k,-}^{in}$, at $\theta$ and $\theta_-$ ($\theta_+ < \theta_-$). The upper panel of Fig. 3 contains calculations of the angular deviations, $\Delta\theta_{+\to-} = \theta_+ - \theta$ and $\Delta\theta_{-\to+} = \theta_- - \theta$ (Fig. 1 clarifies the nomenclature). Note that $\Psi_{k,-}^{in}$ sustains a critical incident angle $\theta_c$ above which the beam scattered at $\theta_-$ vanishes. Next, we will calculate the amplitudes $A_\pm(\mathbf{k},\pm)$, where the labels $\pm$ in the brackets refer to scattered beams and $\pm$ in the subscript denote incident waves. The simplest model of the barrier is described by a step function: $U(x) = U_0$ for $x < 0$ and $U(x) = 0$ elsewhere. By matching the wave function and its flux at $x = 0$, we obtain the amplitudes and the current densities of scattered beams: $T_{--} = |A_-(-)|^2$, $T_{-+} = |A_-(+)|^2$, $T_{++} = |A_+(+)|^2$, and $T_{+-} = |A_+(-)|^2$ (lower panels of Fig. 3). The intensities of scattered beams depend weakly on the barrier height if $U_0 - E_F > \Delta_{SO}$. If $\eta = (\Delta_{so}/E_F)^{1/2} \ll 1$ and $\theta - \pi/2 \gg \eta$, the intensities of scattered beams are given by: $T_{--} = T_{++} = [1 - \cos(2\theta)]/2$ and $T_{-+} = T_{+-} = [1 + \cos(2\theta)]/2$. Figure 3 shows numerical result for the coefficients $T$ with the barrier height $U_0$ at 80 m$e$V. For the incident wave $\Psi_{k,-}^{in}$ at $\theta > \theta_c$ the solution of Schrödinger equation in the region $x > 0$ contains an exponentially decaying contribution, $e^{-\beta x}$. In Fig. 3, $T_{-+}$ at $\theta > \theta_c$ represents a squared amplitude of this exponentially decaying contribution.

We now assume that the incident beam of electrons is not spin-polarized and contains electrons in both $\Psi_{k,+}^{in}$ and $\Psi_{k,-}^{in}$, as is the case in our experiment. Due to the SOI and conservation of the momentum $k_y$, the scattered wave will consist of a triple beam (Fig. 1). The side beams are fully spin-polarized whereas the middle beam does not carry spin polarization. Charge conservation arguments can be used to calculate the current carried by the unpolarized middle beam ($I_0$) and by the polarized side beams ($I_+$ and $I_-$), assuming unit current in the incident beam. For example, if $\theta = 45°$ and $\eta \ll 1$, the currents in the scattered beams in Fig. 1 become $I_0 \approx 1/2$ and $I_+ \approx I_- \approx 1/4$.

To experimentally observe the predicted multi-beam scattering, a geometry yielding large angles $\Delta\theta$ is advantageous. For our parameter values, these angles can reach ~ 10° at incident angles $\theta \approx 70°$ (Fig. 3). Concomitantly however, the intensity of spin-polarized side beams decreases with increasing $\theta$. Hence, in the experiments a compromise was reached with $\theta = 60°$, resulting in an equilateral triangle. The magnetic field exploited to sweep the three beams over the exit aperture is much weaker than the effective SOI field, and hence does not perturb the spin orientation of electrons, and does not invalidate the theoretical treatment above. For the exit aperture width of 0.2 μm, we find theoretically that the average distance between the cutoff magnetic fields is about 3.4 mT, remarkably close to the average difference between minima of about 3 mT observed experimentally (Fig. 2).

In conclusion, we demonstrate experimentally and theoretically spin-polarized reflection off a barrier in an InSb/InAlSb heterostructure. We show that the spin-orbit coupling leads to different reflection angles for different spin polarizations. The spin-polarized beams resulting from the interaction with the barrier can be utilized toward various spin electronics or quantum computational realizations.

We thank G. Khodaparast for informative discussions. J. J. H. acknowledges support from NSF grant DMR-0094055 and M. B. S. from NSF grants DMR-0080054 and DMR-0209371.

FIGURE CAPTIONS

FIG. 1: *Upper left*: schematic of the geometry. Electrons are injected at the upper aperture, scatter from the left barrier and are collected at the lower aperture. A perpendicular magnetic field $B$ allows the trajectories to sweep the lower exit aperture. Trajectories are indicated for $B = 0$ (dotted line) and $B = 7$ mT (solid line). *Lower left*: image of sample $S_1$. *Upper right*: the scattering geometry and the nomenclature for the spin-polarized beams. The spin states are denoted + and -. The deviations from specular reflection, $\Delta\theta_{+\to-}$ and $\Delta\theta_{-\to+}$, lead to spin-polarized reflected beams. *Lower right*: geometrical interpretation of the spin-polarized scattering event, with incident and reflected wave vectors at the Fermi surface (for clarity only scattering of incident + spin states is depicted). Energy and the momentum parallel to the barrier are conserved.

FIG. 2: The four-contact resistance of the triangular structures $S_1$ and $S_2$, versus the perpendicular applied magnetic field $B$. The arrows indicate the values of $B$ where beam cutoffs occur. *Insert*: magnetoresistance of an anti-dot lattice fabricated on the same heterostructure (anti-dot diameter 0.4 μm, periodicity 0.8 μm), showing, for comparison, a featureless negative magnetoresistance background. Geometrical resonances appear at higher $B$ (not shown).

FIG. 3: Calculated spin-dependent angular deviations $\Delta\theta_{-\to+}$ and $\Delta\theta_{+\to-}$ (top panel), and transmission coefficients (lower panels), as a function of the incident angle. Inserts illustrate the kinematics of the spin-dependent reflection. Parameters $\alpha_{SO} \approx 1\times 10^{-6}$ meV cm and $\gamma \approx 10^{-14}$ cm$^2$ were used.

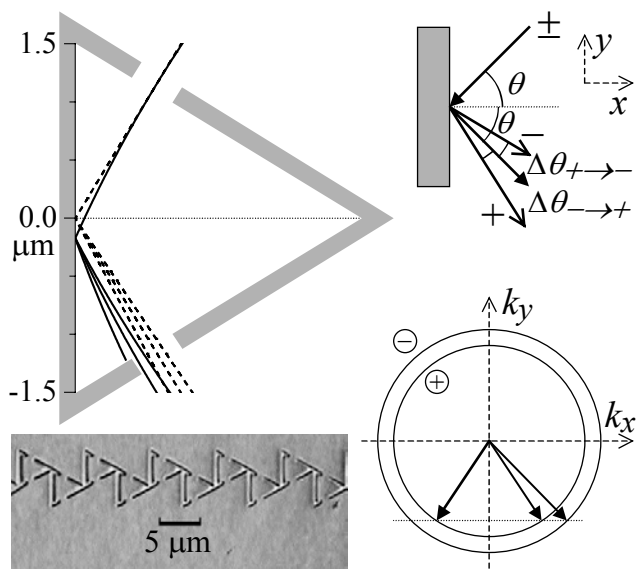

FIGURE 1

(Hong Chen *et al.*)

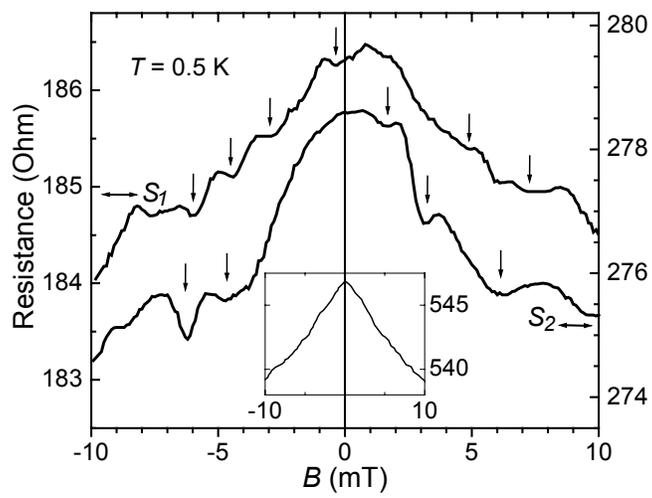

FIGURE 2

(Hong Chen *et al.*)

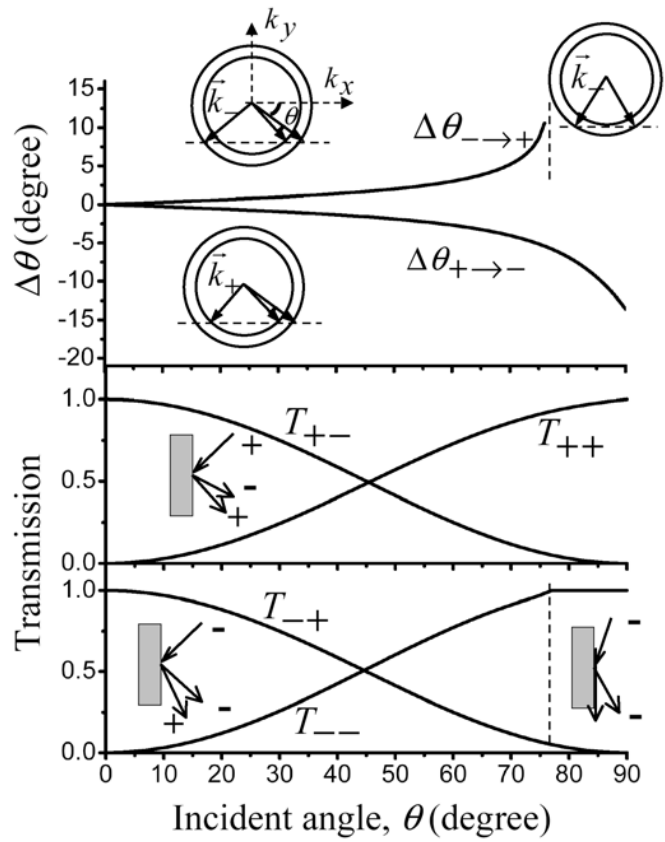

FIGURE 3

(Hong Chen *et al.*)